# Simulations

*N. M. Ngada*
DESY, Hamburg, Germany

**Abstract**
The complexity and cost of building and running high-power electrical systems make the use of simulations unavoidable. The simulations available today provide great understanding about how systems really operate. This paper helps the reader to gain an insight into simulation in the field of power converters for particle accelerators. Starting with the definition and basic principles of simulation, two simulation types, as well as their leading tools, are presented: analog and numerical simulations. Some practical applications of each simulation type are also considered. The final conclusion then summarizes the main important items to keep in mind before opting for a simulation tool or before performing a simulation.

## 1 What is simulation?

Simulation can be defined simply as an abstraction of reality, a representation of real-world activities. An even better definition would be that simulation is a procedure used to analyse physical systems which are too complex for theoretical considerations. However, simulation is performed by developing a model, and in turn a model constructs a framework to describe a physical system. In other words, simulation refers to the result of running a model. This means that one would not 'build a simulation'; rather, one would 'build a model', and then 'run a simulation'.

To illustrate the meaning of simulation, let us take an example: you have to perform some experiments on a physical system, for instance a car that hits a concrete wall at approximately 350 km/h. This experiment can be performed in two ways: either the physical system itself or a model of the physical system can be used. However, if the experiment needs to be repeated more than once a physical system may not be cost effective once cost and time-consuming factors are taken into account.

Using a cheaper physical model (e.g., an old car or other models of car with cheaper materials), or even better a mathematical model of your physical system, can minimize the cost and the time-consuming factors.

The experiment results from the mathematical model can be achieved by using either analytical methods for accurate results or simulation models for approximate results.

Therefore, on the one hand, a good understanding of your physical system will help you to build a good simulation model; on the other hand, a good simulation model will help you to optimize your real system. So, there is somehow a kind of correlation between a real system and its simulation model.

In brief, simulation is no more than a gross simplification of reality, because it includes only a few factors of the physical system. Simulation is only as good as the underlying assumptions. In other words, false assumptions mean false simulation models.

## 2  Why simulation?

There may be several reasons why a simulation is appropriate. One of the main reasons is the fact that simulation allows experiments to be conducted without exposure to risk. In fact, a study of the real system can be too complicated, too expensive, or even too dangerous. Furthermore, simulation can be useful when the real system does not yet exist or is not understood. In addition, it may not be possible to observe the real system directly, or it could be working too fast (e.g., an electrical network) or too slow (e.g., geological processes) to be analysed directly. The last important point is that nowadays the complexity of systems in the field of power converters makes the use of simulation unavoidable.

### 2.1  Fields of application

Simulation is very versatile and suitable for applications in the field of engineering, physics, astrophysics, chemistry, biology, economics, social science, training, education, video games, and more. In this paper, only simulation for engineering applications, especially in the field of power converters for particle accelerators, will be taken into account.

### 2.2  Advantages of simulation

The biggest benefit of simulation is that time and cost are saved during the real-system implementation because designing, building, testing, redesigning, rebuilding, retesting, and so on could be very expensive in both time and money. In addition, simulation provides understanding about how systems really operate without building them. Moreover, simulations are repeatable and can be optimized at any time to give results that are not measurable with current technology.

### 2.3  Disadvantages of simulation

There are, however, some disadvantages of simulation of which the simulator should be aware. For instance, the simulation results could be completely wrong because of a few input data errors. Moreover, sometimes the results of complex simulations are difficult to understand and analyse. Apart from that, simulation cannot solve problems by itself, since a good simulation needs a basic understanding of the real system. Finally, building a good simulation tool can be very time consuming for the model constructor, and purchasing such a tool can be very expensive.

## 3  Principles of simulation

In the field of power converters for particle accelerators, simulation could be split up into two different types: analog simulation and numerical simulation. Which is used depends on the type of experiment that is to be performed.

On the one hand, analog simulation is summarized mathematically by a system of differential (or integral) equations. For this equation to be solved, a model of components and Kirchhoff's circuit laws must be put together. The system of equations is solved by using algebraic and arithmetic methods as the main technique for the entire model simulation.

On the other hand, for numerical simulation, after geometrical representation a mesh is created to divide an object into tiny elements which can be easily studied and recombined for the entire simulation system. Furthermore, the material properties and geometrical boundary conditions must be taken into consideration before solving the problem as a system of differential (or integral) equations. Here the numerical approximation is used as the main method to solve the equation system.

Generally, analog simulation is more appropriate for time-dependent systems, when the time can be precisely controlled. It is mainly used for circuit simulation. Numerical simulation is more suitable for space-dependent systems, where the space can be precisely controlled, and it is mainly appropriate for field simulation. In other words, numerical simulation is useful when the evolution of the real system in space is required, whereas analog simulation is more convenient when the evolution of the real system in time has to be considered. Nevertheless, both analog and numerical simulations require some input data as starting values and some boundary conditions in order to perform the simulation. Both simulation types are independent, but could be used to analyse the same physical system.

Figure 1 shows one field of application for the analog simulation in which the time dependence is important: the current and voltage waveforms for an inductance circuit. Figure 2 illustrates a typical application for the numerical simulation: the magnetic field strength along a vertical cut plane. The result of the simulation is calculated at a certain point in space on the vertical cut plane.

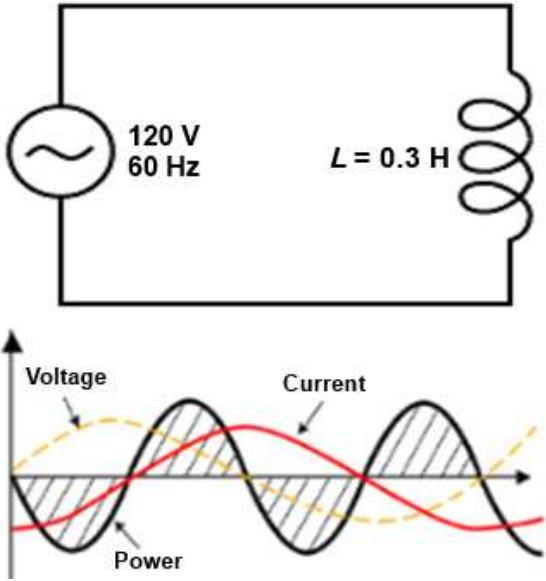

**Fig. 1:** Analog simulation: current and voltage waveforms for a pure inductance circuit

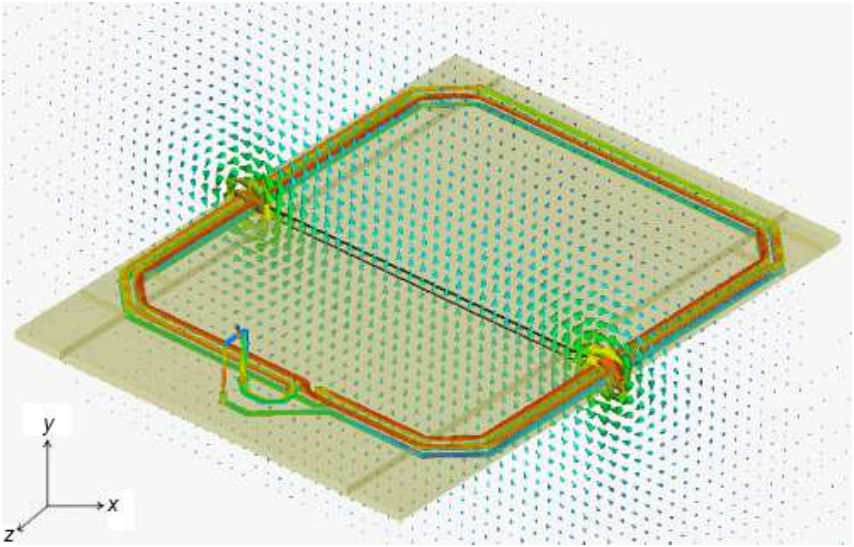

**Fig. 2:** Numerical simulation: surface current distribution of the coil and magnetic field strength along a vertical cut plane.

# 4 Types of simulation

Simulation, as already stated, can in general be split into two different types: analog and numerical. Now let us have a closer look at each type of simulation. Some practical applications of each simulation type will be considered.

## 4.1 Analog simulation

### *4.1.1 Analog simulation tools*

We present some of the most utilized analog simulation tools in the field of power converters for particle accelerators.

#### *4.1.1.1 PSpice (Personal Simulation Program with Integrated Circuit Emphasis)*

PSpice is an analog and digital circuit simulation tool. It started as a simulation tool for low-power electronic circuits and it has been on the market for about 30 years. A large library of PSpice models for various electronic components exists. However, the representation of numerical blocks and controllers is difficult to achieve. The cost of a PSpice licence starts from €7000 for industry and from €3000 for universities. Student licences and demo versions are available, but with limited model sizes.

#### *4.1.1.2 Matlab/Simulink/SimPowerSystems/PLECS*

Matlab—a mathematical tool intended primarily for numerical computing—was first developed more than 40 years ago. However, optional toolboxes such as SimPowerSystems combined with Simulink allow the simulations of electrical power systems including power electronics. A Matlab licence combined with Simulink and SimPowerSystems starts from €8000 for industry and from €2000 for universities. Student licences and demo versions are available for a small amount, but they have limited model sizes. An additional toolbox, which can be combined with Matlab for the simulation of power electronics, is PLECS. This is a fast and reliable power toolbox for Matlab.

#### *4.1.1.3 Simplorer*

ANSYS Simplorer is a multidomain simulation tool for complex power electronic and electrically controlled systems. Simplorer basically integrates four modelling techniques (e.g., digital simulator, circuit simulator, block diagram, and tate machine) that can be used concurrently within the same schematic. Simplorer can be interfaced to many other simulation tools. A Simplorer licence starts from €3500 for universities. Student licences and demo versions are available, but with limited model sizes.

#### *4.1.1.4 PSIM (PowerSim)*

PSIM is one of the simulators that was developed 20 years ago specifically for power electronics, but it can be used to simulate any electronic circuit. PSIM is one of the fastest simulators for power electronics simulation, and therefore it is optimized for the tasks that arise in this field. This results in a faster simulation time. PSIM can be interfaced to Matlab/Simulink in order to use the full mathematical power of Matlab. A PSIM licence starts from €1700 for Industry and from €280 for universities. Student licences and demo versions are also available for a small amount, but with limited model sizes.

#### *4.1.1.5 LTspice IV*

LTspice IV is a freeware tool for analog circuit simulation, produced by Linear Technology Corporation. LTspice IV is considered as one of the best freeware tools available for circuit simulation. It started as a simulation tool for models to ease the simulation of switching regulators, but other models of components have been added for electrical circuits. LTspice IV is the most widely distributed and utilized SPICE (Simulation Program with Integrated Circuit Emphasis) program in the industry.

*4.1.1.6   CASPOC*

This tool is designed for the simulation of power electronics and electrical drives. CASPOC is used in the design and simulation of complex power and control devices and systems. It is appropriate for multiphysics (e.g., Computational Fluid Dynamics, mechanical, thermal, or electromagnetic) control systems. CASPOC is the only simulator on the market with a circuit animation feature, which contains a 'freeze and go back' function. A freeware version of CASPOC is available; prices and conditions for industry or universities are unknown to the author.

*4.1.2   Field of application 1: temperature simulation for the European XFEL at DESY Hamburg*

At DESY Hamburg, the European X-ray Free-Electron Laser (XFEL) linear accelerator is currently under construction. Figure 3 shows the cross-sectional view of the tunnel for the particle acceleration part, the so-called XTL tunnel. Figure 4 presents the longitudinal view of the underground XFEL tunnels, with a total length of about 3.4 km, a depth from 6 m up to 38 m, and diameters of 5.2 m for the XTL tunnel and 4.5 m for the photon tunnels. Simulations were very helpful in fixing the required temperature profile for the XTL tunnel.

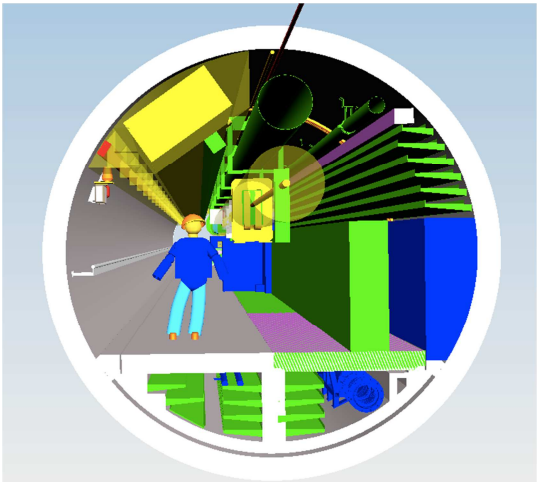

**Fig. 3:** Cross-sectional view of the 2.1 km XTL tunnel

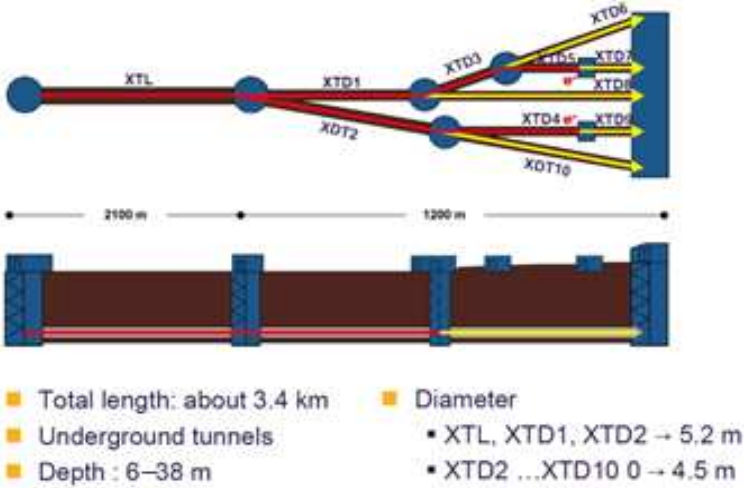

**Fig. 4:** Longitudinal view of the XFEL tunnels

*4.1.2.1 Motivation*

Many LLRF (Low Level Radio Frequency) signal cables whose transmission time is very sensitive to the temperature variation, are installed in the XTL tunnel. For a high-precision machine like XFEL, such effects are undesirable. In order to prevent the changes in propagation time of the LLRF signals, a stable temperature profile along the tunnel is required. This sensitive stability allows only a temperature variation in the range of ±1 K, even during different operational modes of the accelerator (e.g., maintenance days, limited access, or full operating time). Since the system does not yet exist and furthermore is very complex to analyse, using simulations was the best way to fix this special temperature requirement concerning the XTL tunnel.

Some simulations had already been done by means of Matlab for the steady-state temperature calculation. In addition, it should have been possible to perform transient analyses with a numerical simulation tool such as ANSYS CFX, but these cost too much in computing time and capacity due to the limited ANSYS CFX licences at DESY.

To achieve this goal despite this fact, the ANSYS Simplorer package was chosen for analog simulation. Two reasons motivated this choice. First, ANSYS Simplorer can handle complex multiphysics circuit systems with transient behaviour (e.g., electrical, thermal, electromechanical, electromagnetic, and/or hydraulic) quite easily. Second, it has a very stable simulation algorithm. Also, enough user licences were available in our department.

*4.1.2.2 Proceeding*

As a first step, several input parameters, such as the heat sources (e.g., lights, cables, waveguides, hot water pipes, pulse transformers, and matching networks and magnets) as well as the heat sinks (e.g., cold water pipes, the fact that the tunnel is underground) were investigated and the inlet temperature defined. In addition, knowledge of the geology of the ground, and previous experience with and temperature measurements in the former accelerator machine HERA, were helpful in understanding the real system and implementing a good and reliable simulation model.

The second step was to transform the physical components of the tunnel (e.g., heat sources, tunnel wall, or air) into the thermal circuit component (e.g., capacitance or resistance) model, and then to add the material properties to these components. However, some heat sources and heat sinks changed every 50 m along the accelerator tunnel. This meant that a new component model was necessary along every 50 m section in the tunnel to complete the entire simulation, as presented in Fig. 5. In this way the entire XTL tunnel was divided into several 50 m sections and was drawn as a model of a thermal circuit in the ANSYS Simplorer schematic. To simulate the transient temperature behaviour in the entire XTL tunnel, a total of 43 models were added in series, as shown in Fig. 6.

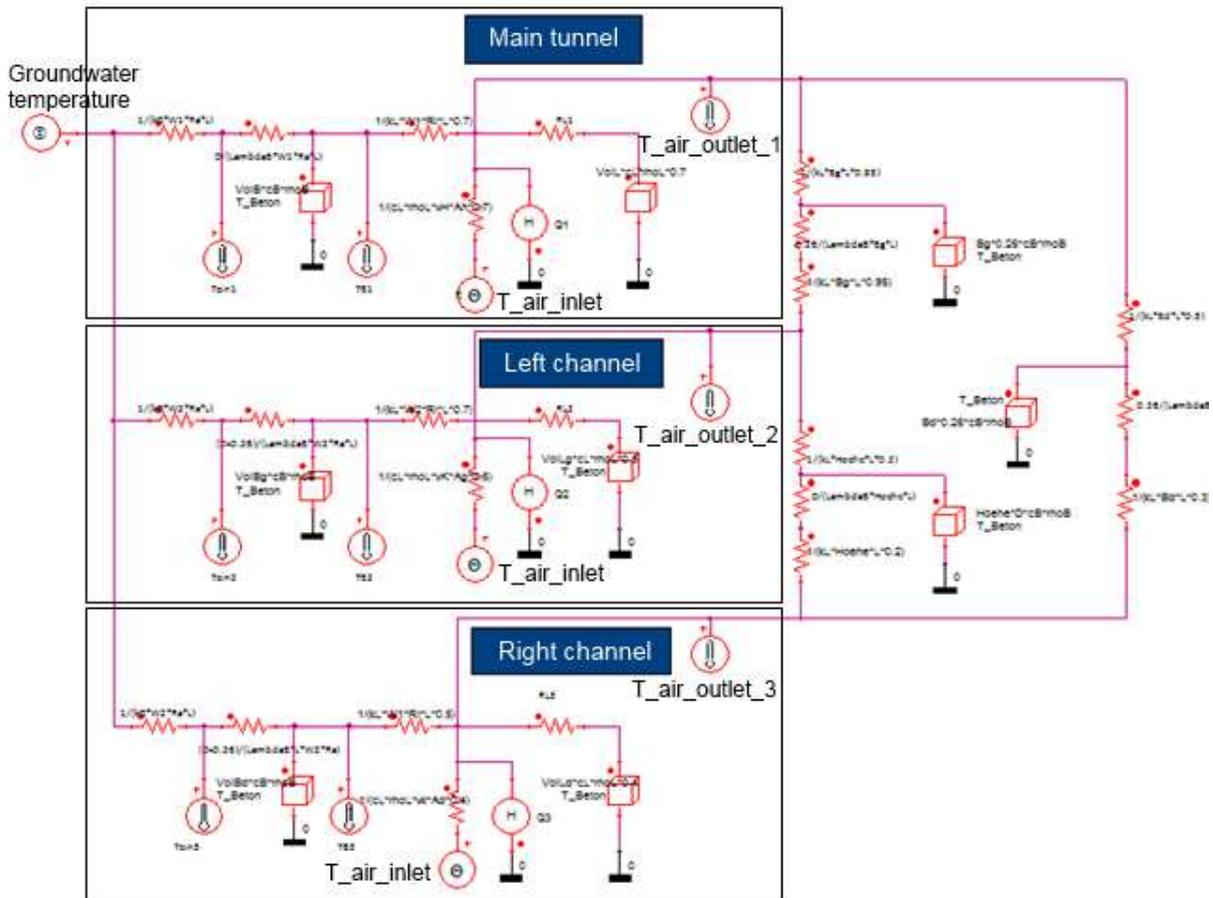

**Fig. 5:** Model of the 50 m XTL tunnel section as a thermal circuit in the schematic of ANSYS Simplorer

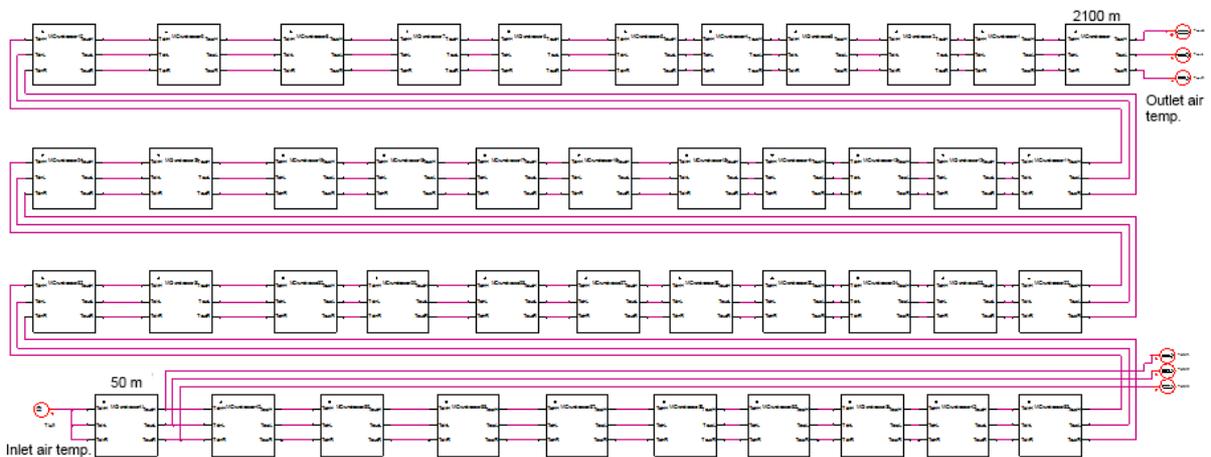

**Fig. 6:** The entire 2.1 km XTL tunnel with a total of 43 model components as a thermal circuit in the schematic of ANSYS Simplorer.

### *4.1.3    Simulation results*

Figure 7 shows a photograph of the empty XTL tunnel at DESY with the lights as the only heat sources. After running the first simulation, several measurements were made in the empty physical tunnel to assist with further adjustments on the simulation model in order to obtain—on 12 March, 2013—the first plot of the temperature profile (see Fig. 8). This figure shows the measurement as well as the simulation for the steady-state temperature behaviour in the XTL tunnel. The underground temperature around the tunnel during the measurement was about 11°C. This achievement is the outcome of about 6 months of simulation analyses as well as measurements in the empty XTL tunnel.

Therefore, the best way to fit the simulation results with the measurements is a good understanding of the real system. Furthermore, some measurements on the real system are necessary in order to perform the appropriate readjustments of the simulation model.

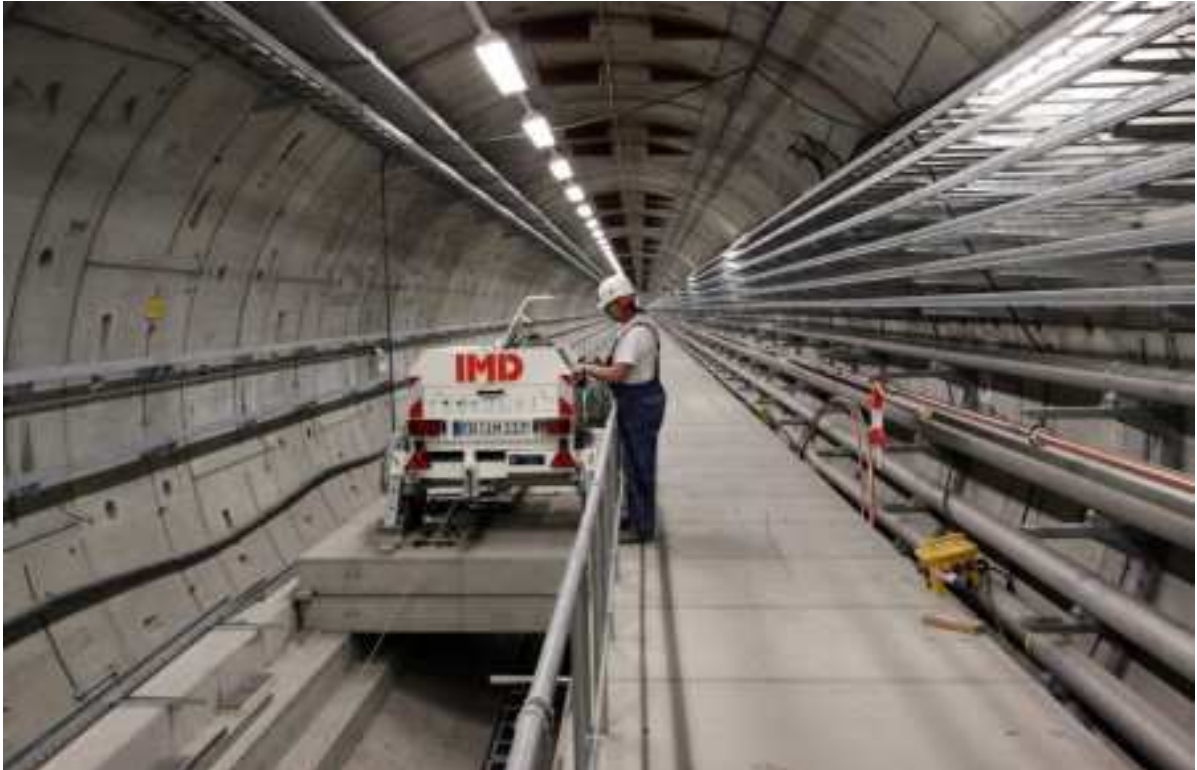

**Fig. 7:** The empty XTL tunnel at DESY Hamburg with the lights as the only heat sources

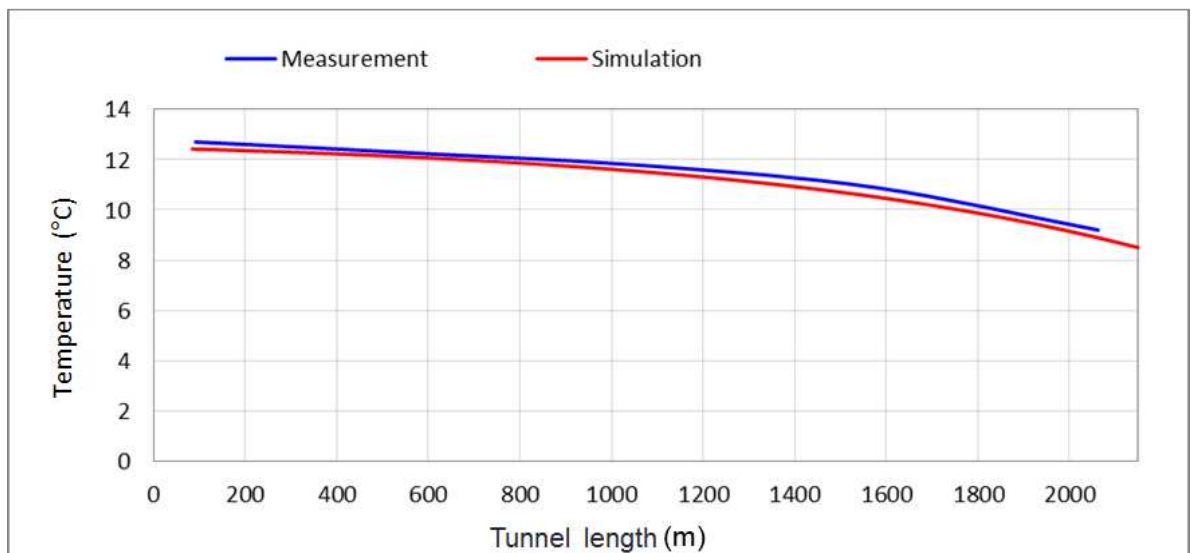

**Fig. 8:** Measurement of the temperature profile in the 2.1 km empty XTL tunnel on 12 March, 2013, compared with the simulation.

Since the simulation of the steady-state temperature behaviour in the empty XTL tunnel was satisfactory, the next step was to analyse the transient temperature behaviour once the tunnel was no longer empty and during different operating modes of the XFEL accelerator machine. Table 1 displays how the most important heat sources are distributed in the three parts of the XTL tunnel: the main tunnel, the left cable channel, and the right cable channel. Table 1 also shows when the heat sources are in service (ON) or not (OFF) for two different operating modes of the machine: maintenance days and the full operating time. The goal now is to simulate the expected temperature profile along the XTL tunnel after a typical service day of about 10 h or after 10 days of full operating time.

Note that the air flows from the end (at about 2.1 km) to the beginning of the tunnel, so that the inlet temperature starts at the end of the tunnel. An inlet temperature of 23°C is chosen for all further temperature simulations.

**Table 1**: Distribution of the most important heat sources in the three parts of the XTL tunnel: the main tunnel, the left channel, and the right channel.

| Heat sources in the XTL tunnel | Tunnel part | Operating time | Maintenance day |
|---|---|---|---|
| Pulse cables, left | Left channel | ON | OFF |
| Pulse cables, right | Right channel | ON | OFF |
| Medium voltage power cables | Main tunnel | ON | ON |
| Low voltage power cables | Left channel | ON | ON |
| Direct current power cables | Left channel | ON | OFF |
| Transformers | Main tunnel | ON | OFF |
| Impedance matching network | Main tunnel | ON | OFF |
| Magnets | Main tunnel | ON | OFF |
| 30°C water pipe 1 (feed line) | Main tunnel | ON | ON |
| 40°C water pipe 1 (outlet flow) | Main tunnel | ON | ON |
| 20°C water pipe 2 ( feed line ) | Main tunnel | ON | ON |
| 25°C water pipe 2 ( outlet flow ) | Main tunnel | ON | ON |
| 20°C water pipe 3 ( feed line ) | Main tunnel | ON | ON |
| 20°C water pipe 4 ( feed line ) | Right channel | ON | ON |
| 20°C water pipe 5 ( feed line ) | Left channel | ON | ON |
| Electronic racks | Main tunnel | ON | ON |
| Waveguides | Main tunnel | ON | OFF |
| Light | Main tunnel | OFF | ON |

Figure 9 shows the temperature profile over the time 50 m from the end of the XTL tunnel during a maintenance day. There are two points to note from this figure: first, the steady-state temperatures are reached after about a day of machine operating with values more or less the same as the inlet temperature. Second, after 10 days of full operation, the temperature values in the three tunnel parts change minimally—about 1.2°C when the accelerator is switched off for 10 h of maintenance. When the accelerator is switched on again for the full operating mode, the temperatures take more than a day to reach the steady-state values.

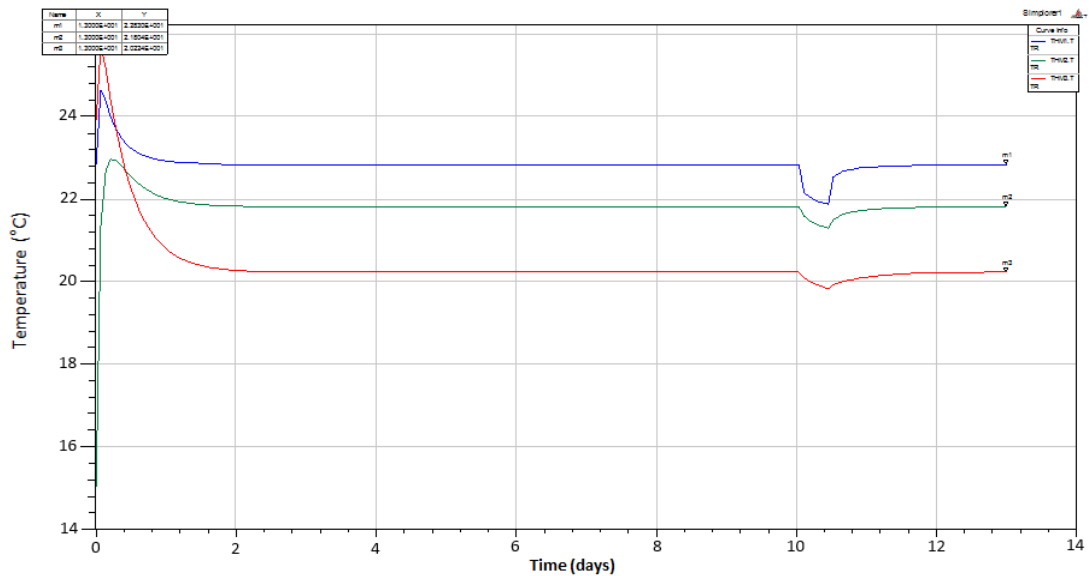

**Fig. 9:** The changes in temperature over time 50 m from the end of the the XTL tunnel (at about 2.1 km). Blue curve: temperature in the main tunnel; green curve: temperature in the left cables channel; red curve: temperature in the right cables channel.

The air became much warmer at the beginning of the tunnel (see Fig. 10), at about 2.1 km away from the position in Fig. 9. Two points should be also emphasized. First, the steady-state temperatures are reached after more than one day of machine operating at approximately 4°C above the inlet temperature. Second, after 10 days of full operation, the temperature values in the three tunnel parts drop by about 7°C after 10 h of maintenance. When the accelerator is switched on again for the full operating mode, the temperatures take more than two days to reach steady-state values.

In this manner the analog simulation provides understanding about how the XTL tunnel really operates. Moreover it was possible using the simulation to consider all the interactions between the heat sources and the heat sinks and finally determine the requirements for stable temperature behaviour in the tunnel during different machine operating modes.

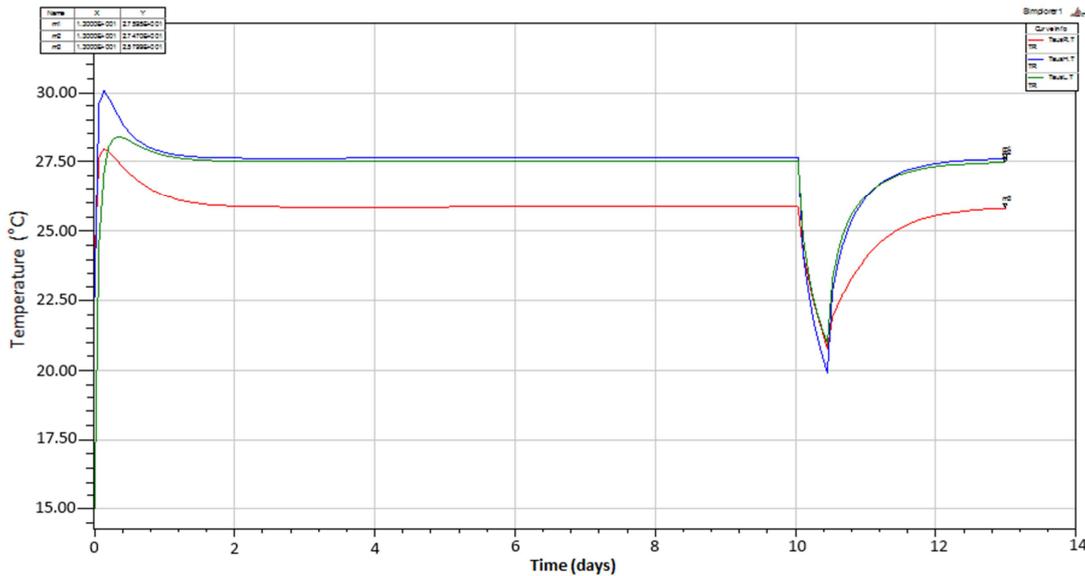

**Fig. 10:** The changes in temperature over time at about 2.1 km from the end of the XTL. Blue curve: temperature in the main tunnel; green curve: temperature in the left cables channel; red curve: temperature in the right cables channel.

## *4.1.4 Round up*

The example shows that for simulation at least a basic understanding of the physical system is required. However, the simulation model should be as simple as possible, but should be complex enough to answer the questions asked. When possible, making measurements on the real system helps to optimize the model for accurate simulation results.

## 4.2 Numerical simulation

### *4.2.1 Numerical simulation methods*

Numerical approximation methods are used to find the solution of numerical simulation problems. Some of the most used numerical methods are discussed in the following.

#### *4.2.1.1 The finite difference method*

The finite difference method (FDM) is the method used to approximate the solutions to differential equations using finite difference equations to approximate derivatives. The principle consists in approximating the differential operators by replacing the derivatives in the equations using differential quotients. The domain is partitioned in space and time, and approximations of the solutions are computed at the space or time points.

#### *4.2.1.2 The moment method*

The method of moments (MoM) is a numerical computational method of solving linear partial differential equations which have been formulated as integral equations. It can be applied in many areas of engineering and science, including fluid mechanics, acoustics, electromagnetics, fracture mechanics, and plasticity.

#### *4.2.1.3 The finite element method*

The finite element method (FEM) is used to find approximate solutions of partial differential equations (PDEs) and integral equations. The solution approach is based on either eliminating the time derivatives completely (steady-state problems) or rendering the PDE into an equivalent ordinary differential equation, which is then solved using standard techniques such as finite differences, etc. In solving PDEs, the primary challenge is to create an equation which approximates the equation to be studied, but which is numerically stable. The FEM is a good choice for solving PDEs over complex domains or when the desired precision varies over the entire domain.

#### *4.2.1.4 The Monte Carlo method*

The Monte Carlo method is based on repeated random sampling to obtain numerical results. The simulation typically runs many times over in order to obtain the distribution of an unknown probabilistic entity. The Monte Carlo method is often used in physical and mathematical problems, and is most useful when it is difficult or impossible to obtain a closed-form expression, or infeasible to apply a deterministic algorithm. Monte Carlo methods are mainly used in three distinct problem classes: optimization, numerical integration, and generation of draws from a probability distribution.

#### *4.2.1.5 The method of lines*

The method of lines is a general technique for solving PDEs by typically using finite difference relationships for the spatial derivatives and ordinary differential equations for the time derivatives.

### 4.2.2 *Numerical simulation tools*

The following tools are among the best known numerical simulation packages in the application field of power converters for particle accelerators.

#### 4.2.2.1 *Quickfield*

Quickfield is based on finite element analysis and is developed and distributed by Tera Analysis Ltd. It is available as a commercial program or as a free program with limited functionality. Quickfield is a stable and fast package, which is mainly used for the simulation of electromagnetic fields. A licence starts from about €1200, depending on the version and the type of application. Therefore it is less expensive for research institutes.

#### 4.2.2.2 *ANSYS CFX*

ANSYS Computational Fluid Dynamics (CFX) simulation software allows the simulation of fluid flow in a variety of applications. It is based on finite element analysis, which was developed by Dr John Swanson. His company, founded in 1970—SASI (Swanson Analysis Systems Inc.)—developed the first versions of ANSYS up to version 5.1. After the sale of the company in 1994 it was renamed ANSYS Inc. ANSYS CFX solutions are fully integrated into the ANSYS Workbench platform. Workbench integrates workflow needs as well as multiphysics functionality (fluid–structure interaction, electronic–fluid coupling, etc.). A student licence with reduced model sizes is available. Price and conditions for industry or research institutes are unknown to the author.

#### 4.2.2.3 *ANSYS HFSS*

The HFSS (High Frequency Structure Simulator) is a finite element method tool for three-dimensional full-wave electromagnetic field simulation from ANSYS and is essential for the design of high-frequency component design (e.g., antenna design). Previously known as the Ansoft HFSS, Ansoft was later acquired by ANSYS. A licence starts from €20,800 for industry and from €14,000 for universities. A student licence with reduced model sizes is less expensive.

#### 4.2.2.4 *ANSYS MAXWELL 2D*

Maxwell 2D is an electromagnetic simulation software program used to develop accurate virtual prototypes of electric machines, actuators, transformers, sensors, and other electromagnetic devices that can be represented in two dimensions. A student licence with reduced model sizes is available. Prices and conditions for industry or research institutes are unknown to the author.

#### 4.2.2.5 *FEKO*

FEKO is a MoM tool developed by EM Software & Systems – S.A. (Pty) Ltd for electromagnetic simulation. FEKO has an online simulation service with the cost of usage per hour per core used, which includes FEKO licence fees.

#### 4.2.2.6 *CONCEPT-II*

The CONCEPT-II software is an advanced electromagnetic field simulator for the numerical computation of radiation and scattering problems in the frequency domain. The code is based on the MoM and integral equations for the electric and magnetic fields. CONCEPT-II is developed by the Institute of Electromagnetic Theory at the Technical University of Hamburg–Harburg (TUHH) and can be used free of charge in academia. Price and conditions for industry are unknown to the author.

### 4.2.3 Field of application 2: grounding of HV power suppliers (modulators) for RF stations for the XFEL

Let's consider now an example of application for the numerical simulation. The European X-ray Free-Electron Laser (XFEL) linear accelerator at DESY Hamburg requires 29 RF stations capable of 10 MW RF power each for electron acceleration in the XTL section. The RF power for the XFEL linear acceleration is generated by klystrons which are installed in the underground XTL tunnel, close to the accelerator modules. The klystrons are powered by HV pulses from modulators which are located in the modulator hall (XHM), above ground on the DESY site. Each modulator is connected to the pulse transformer at the klystron by means of a long triaxial cable and pulses up to 12 kV and 2 kA with a duration of 1.7 ms and a nominal repetition rate of 10 Hz. The grounding simulations of these modulators assist in greater understanding of some EMC effects occurring during commissioning.

Figure 11 shows a combination of devices connected from the transformer to the modulator. HV racks distribute the power among the 29 modulators and the HV racks. The modulators are interconnected by a system of protection earth, grounding, and power cables, which are three-phase cables without a neutral conductor.

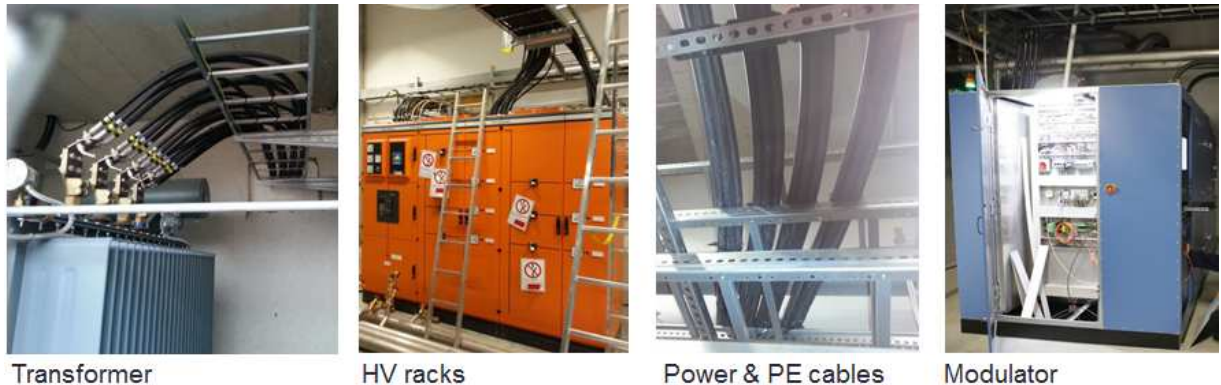

**Fig. 11:** Combination of devices comprising the modulator system in the hall

#### 4.2.3.1 Motivation

Measurements taken after installing and commissioning the first modulator showed some electromagnetic interference which was not understood. The sources of a 50 Hz high current of more than 50 A peak to peak, measured on the PE conductor were unknown. Faults in the insulation of the power cables and PE conductor had already been excluded after insulation measurements were made. The main suspect was the EMI current induced from the power cables.

The goal was to perform simulations to confirm the suspicions and finally to find a solution for the optimization of the grounding system of the 29 modulators. To achieve this, Quickfield was chosen from the tools for numerical simulation for several reasons. Quickfield is very easy to learn and offers the useful possibility of combining electrical circuits with field simulations. In other words, geometrical models from the numerical simulation could be easily transformed as electrical circuits in the same schematic of Quickfield. Moreover it is a multiphysics tool—very stable and very fast with various analysis types (e.g., a.c., d.c., and transient electromagnetics, electrostatics, steady-state and transient heat transfer, stress analysis). The main disadvantage is that only basic components for electrical circuit analysis (e.g., resistors, capacitors, inductors, diodes, voltage sources, and current sources) are available in the library of Quickfield components.

#### 4.2.3.2 Technical data for the modulator

The technical data for the modulator is given in Table 2 and Fig. 12 shows the overview of the modulator hall at DESY Hamburg.

**Table 2:** Technical data for the modulator

| | |
|---|---|
| Number of modulators | 29 |
| Output voltage | 0–12 kV |
| Output current | 0–2 kA |
| Average output power | max. 380 kW |
| Maximum pulse power | 16.8 MW |
| Pulse duration | 0.2–1.7 ms |
| Pulse repetition rate | 1–30 Hz |

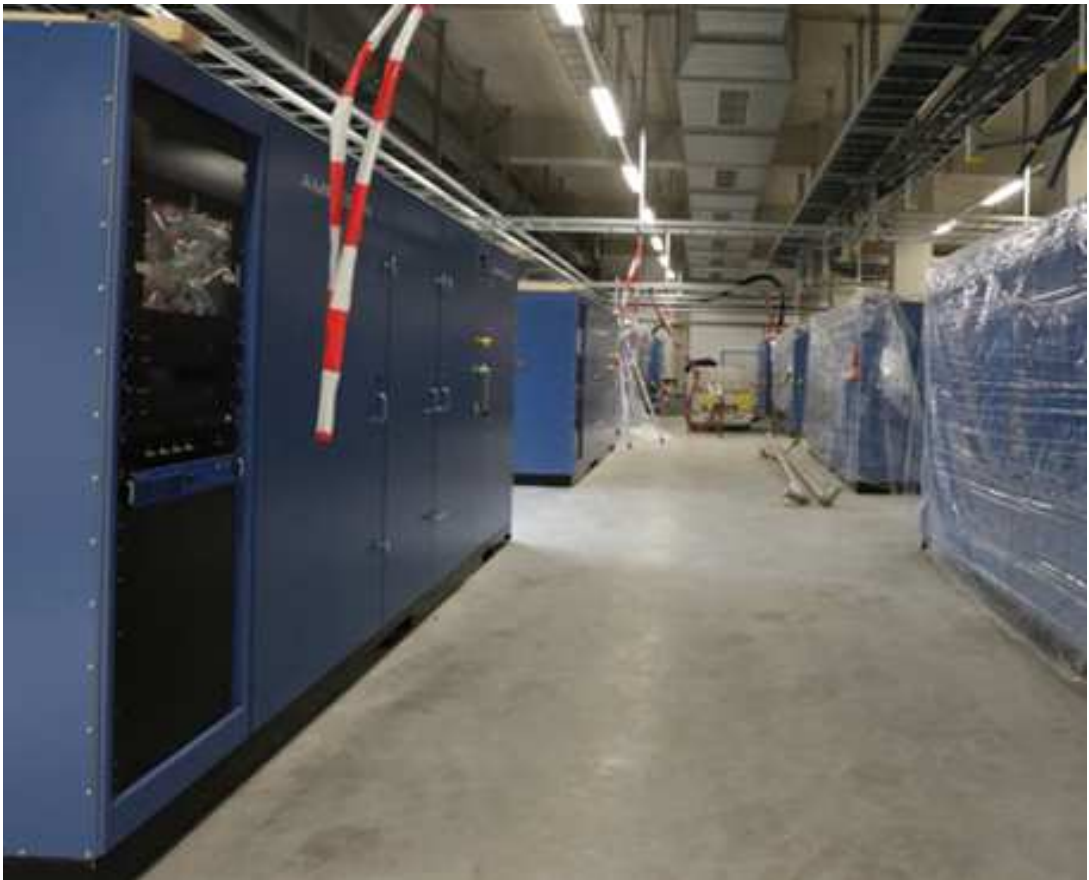

**Fig. 12:** Overview of the modulator hall

*4.2.3.3 Proceeding*

Figure 13 shows an overview of the grounding schematic of the modulators connected as a TNS (Terre Neutre Séparé) system, with PE (protective earth) and N (Neutral) conductor separated from the transformer grounding point. The neutral conductor between the HV racks and the modulators is not required because of the symmetrical properties of the modulators as loads. The cable length between the modulator and the HV racks is about 30 m. The schematic of Fig. 13 was translated as a space-dependent system in Quickfield for field simulation analysis. The field simulation was used to visualize the influence of the electromagnetic field propagation around the power cables. First, the geometry of the power cable systems and surrounding devices (e.g., cable trays, PE conductors) was drawn to build a model for the numerical simulation that is as accurate as possible. Then the geometrical models were transformed into electrical circuit components for further electrical circuit analysis.

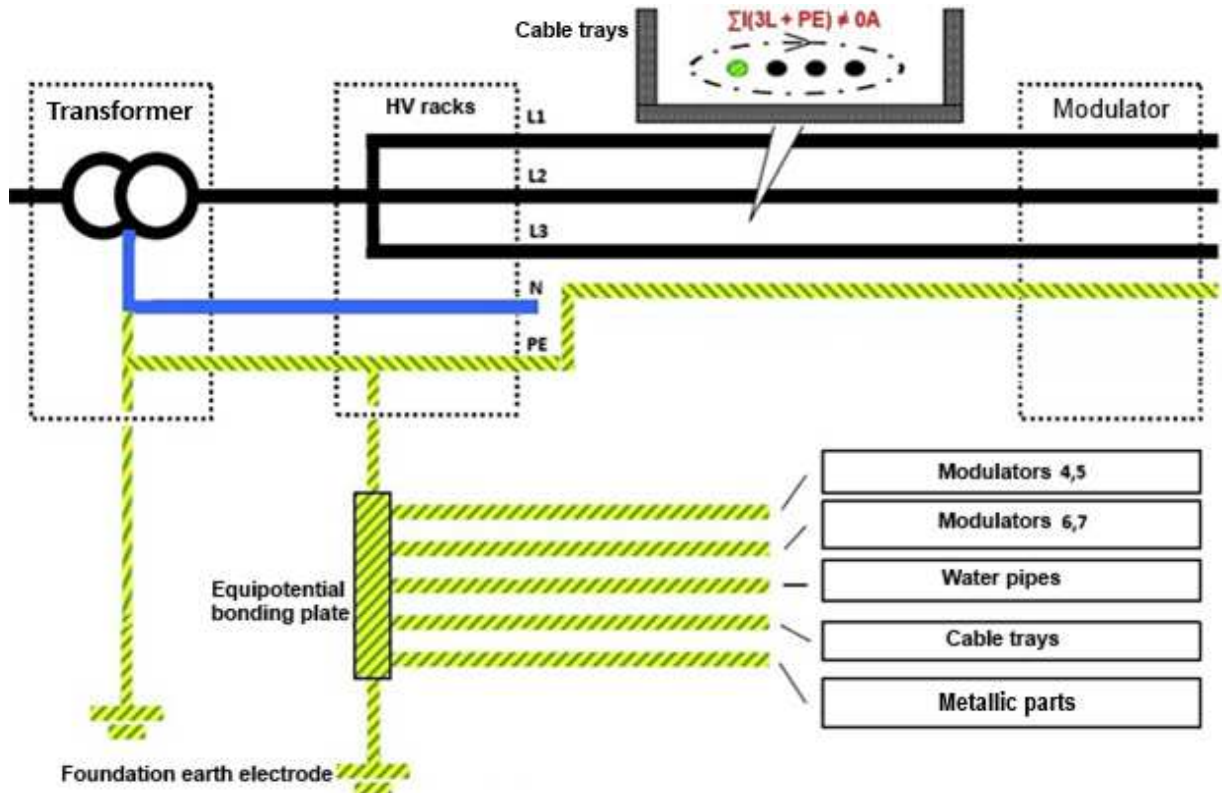

**Fig. 13:** Overview of the grounding schematic of the modulators

### *4.2.4   Simulation results*

Figures 14–19 show the simulation results in three different cases: when the PE conductor is near to, between, or about 25 cm away from the power cables. These results will then be compared with the measurements.

#### *4.2.4.1   PE conductor near to one of the three power cables*

Figure 14 (left) shows the magnetic field and the current density field simulation. The electrical circuit schematic from the field simulation can be observed on the right. Fig. 15 illustrates the measurement for simulations results verification. The effective (rms) of the current on the PE conductor between both figures is almost the same. This means the good accuracy of the simulation model compared to the real system. The simulation analysis demonstrates that the power conductors induce uncompensated annoying currents on the PE conductor and the cable trays around. The inducted current on the PE conductor is up to 10% of the power conductors current, which is not desirable for a high precisely accelerator machine like the European XFEL. Then the high PE conductor current could flow through the ground and finally disturb the electrons acceleration in the machine. Furthermore that current could interfere as noise signal for the machine beam monitoring. Further simulations will show how to reduce the inducted current in the PE conductor.

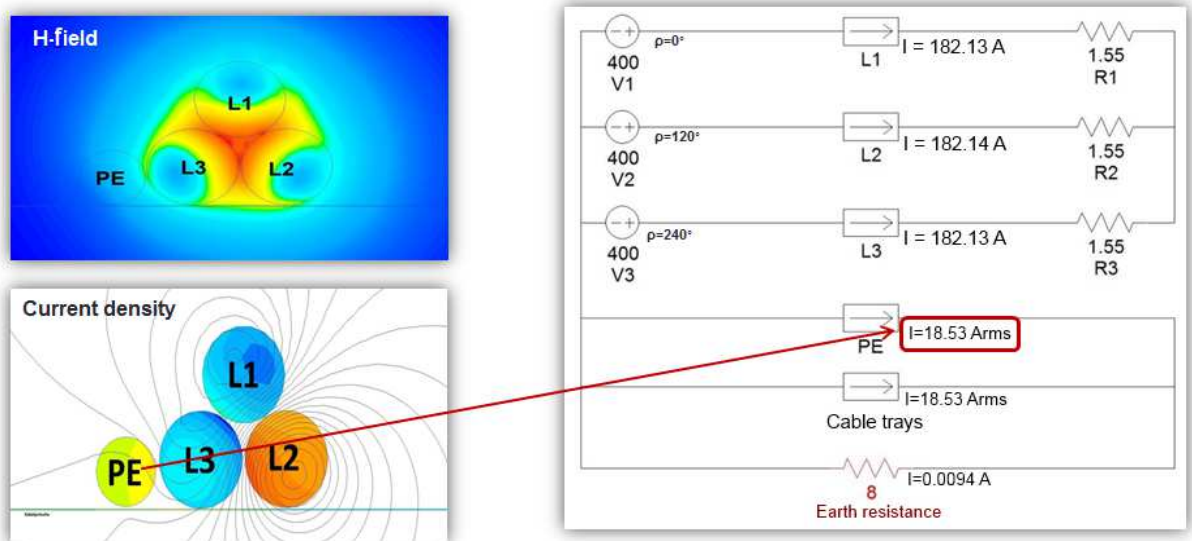

**Fig. 14:** Simulation with the PE conductor near to one of the three power cables

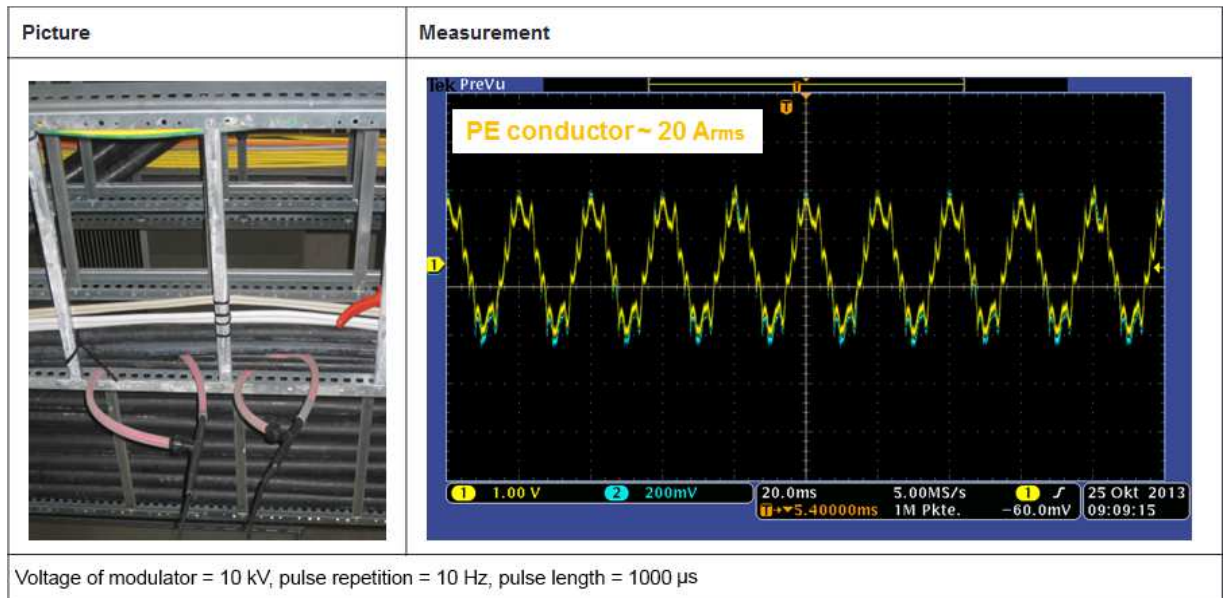

**Fig. 15:** Measurement with the PE conductor near to one of the three power cables

*4.2.4.2   PE conductor between the power cables*

Figure 16 illustrates the results of simulation when the PE conductor is between the three power cables. The current on the PE conductor is greatly reduced, from about 20 $A_{pp}$ to about 6 $A_{pp}$, because the electromagnetic fields between the power cables cancel each other out. The H-field and the current density simulation in Fig. 16, as well as the measurement shown in Fig. 17, confirm this. The difference in value of the PE conductor current in the simulation and that in the measurement is because it is difficult to fix the PE conductor exactly in the middle of the three power cables. The blue sine wave in Fig. 17 represents the uncompensated current measured around the three-phase power cables.

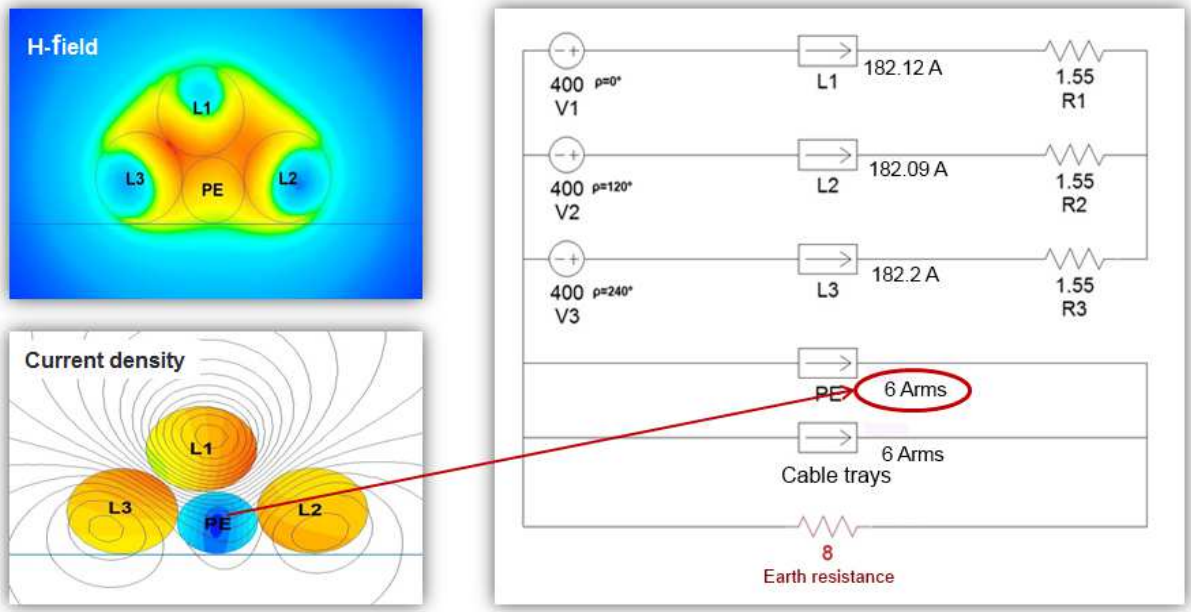

**Fig. 16:** Simulation with the PE conductor between the power cables

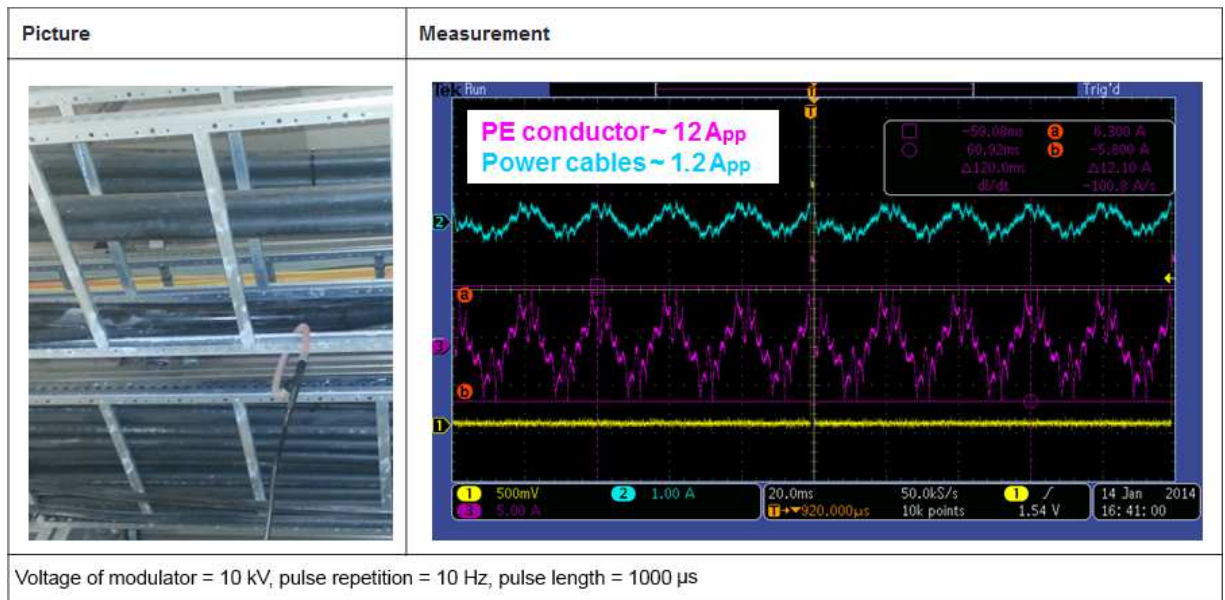

**Fig. 17:** Measurement with the PE conductor between the power cables

### 4.2.4.3   *PE conductor about 25 cm away from the power cables*

Figure 18 shows the results of the simulation when the PE conductor is fixed on the cable tray about 25 cm away from the three power cables. The current in the PE conductor is now greatly reduced, to about 2.8 $A_{pp}$ because the strength of the electromagnetic field decreases further away from the power cables. The simulation results in Fig. 18, as well as the measurement in Fig. 19, demonstrate this. The blue sine wave in Fig. 19 represents the uncompensated current measured around the three-phase power lines.

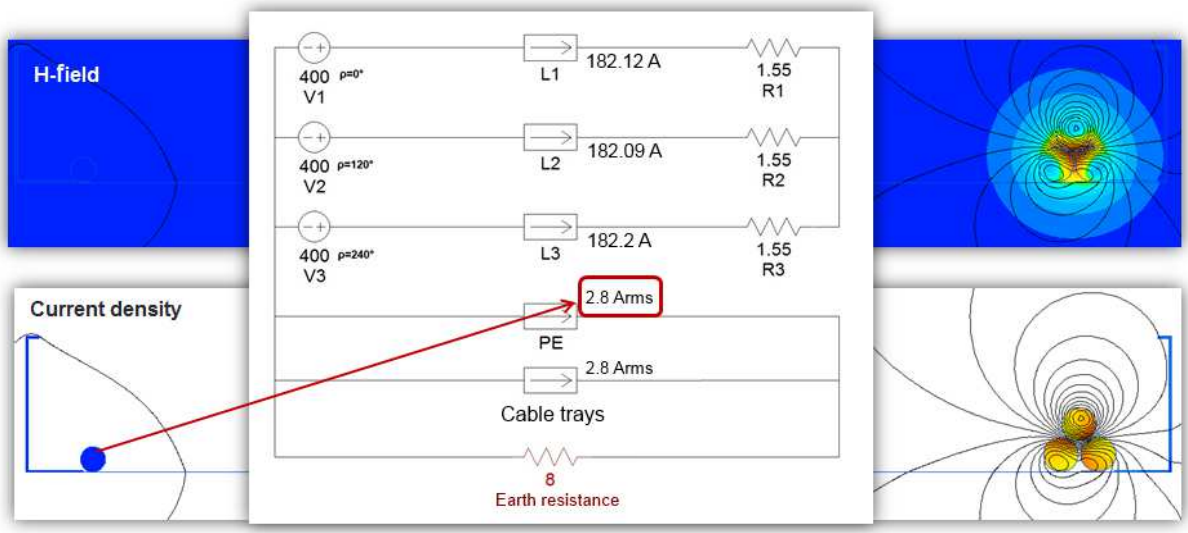

**Fig. 18:** Simulation with the PE conductor about 25 m away from the power cables

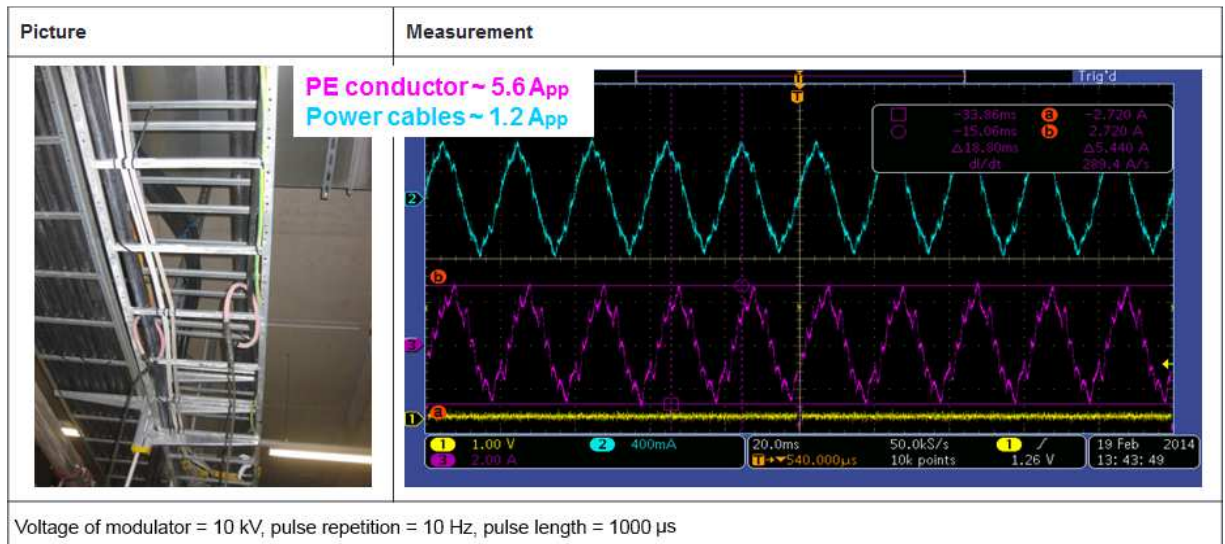

**Fig. 19:** Measurement with the PE conductor about 25 cm away from the power cables

### 4.2.5   Roundup

Simulations, supported by measurements, help in finding a advantageous grounding system for the modulators by allowing a choice of cables with better EMI rejection properties as well as optimizing the cabling for the grounding. In addition, simulations eliminate the costs of making modifications after installation.

## 4.3   Conclusion

### 4.3.1   Challenges in the world of simulation

Despite the huge improvements realized by building simulation tools in over the past few years, further improvements are welcome. Engineers can probably expand this list, but the following improvements in the world of simulation are still desirable for some simulation packages:

   a)  more intuitive software design to make the usage even easier;

b) faster models for lower simulation time;

c) models and results transfer between different simulation tools and operating systems;

d) better user support and extended online help;

e) lower licence costs.

### 4.3.2 *Expectations for a good simulation tool*

Some basic expectations for a good simulation tool comprise:

a) comfortable and intuitive schematic design;

b) easy interpretation of the error messages;

c) robust execution of the simulation;

d) simulation result formats which can be exported to other programs for further analysis;

e) good user support from the manufacturer;

f) portability of models from one program version to the following ones.

### 4.3.3 *Checklist before opting for a simulation tool*

There are many powerful simulation tools available, all of which have some advantages and disadvantages. A few guidelines useful for selecting a simulation tool that will meet your needs are listed in the following.

1) Before expending any effort researching simulation tools, the organization should commit to investing both the necessary money and staff time into purchasing and learning how to use a simulation software program. Depending on the type of simulation tool selected, the price for a single licence can be very expensive.

2) Perhaps the most important step in selecting simulation software is to state clearly the problem (or class of problems) that you would like to address. This must include a general statement about what you would like the simulation tool to do.

3) Because simulation is such a powerful tool, a wide variety of approaches and tools exist to assist in understanding complex systems and to support decision making. Before trying to survey all the available tools, you must first decide upon the general type of tool that you require (e.g., analog or numerical simulation).

4) This step involves developing a set of functional requirements that you would like the software tool to have. Note that requirements specify what the simulation software will do, not how. They should be as concise as possible (e.g., should be able to support a Monte Carlo simulation, a.c., d.c., transient, etc.).

5) An evaluation version of each product should eventually be obtained to test the software. Although this is necessary, it can be time consuming, since there will be a learning curve associated with each product.

### 4.3.4 *Important points to achieve accurate simulation results*

Regardless of the chosen type of simulation, accurate output results always depend on the following.

1) The simulation model should be as simple as possible—complex enough only to answer the questions asked.

2) A basic understanding of the real system is necessary. The expectation that the simulation tool will be 100% correct is wrong, and a lot of time can be spent realizing that.

3) The interrelation between the simulation model and the real system will help you to build an accurate model and to determine if the simulation results fit the physical system.

A famous quotation states: "Those who can, do. Those who cannot, simulate." But nowadays the complexity of power systems makes the basic understanding of the real system before simulation indispensable. Simulation does not replace understanding.

**Acknowledgements**

I wish to thank H.J. Eckoldt for the helpful discussions we had and for the enlightening comments on the present topic. My thanks also go to all my MKK colleagues, who have encouraged me after presenting this topic.

**References**

[1] J. Banks, Simulation in practice, AGIFORS 2002, Rome (2002).

[2] F. Jenni, Simulation tools, CAS – power converters for particle accelerators, Warrington, United Kingdom, CERN 04-05 (2004).

[3] J. Banks, J. Carson, B. Nelson, D. Nicol, Discrete-event system simulation, $5^{th}$ edition, Prentice Hall, Upper Saddle River, 2010.

[4] Prof. Dr.-Ing A. F. X. Welsch, Praktikum für die numerische Feldberechnung mit der Finite-Elemente-Methode, Fachbereich Elektro- und Informationstechnik, May 2007.

[5] GoldSim Technology Group, Selecting simulation software (2009).

[6] P. Fischer, VLSI Design WS07/08 – Analog simulation, TI, University of Mannheim, 2007.

[7] M. B. Cutlip and M. Shacham, The numerical method of lines for partial differential equations. Polymath paper,1998

**URLs**

http://www.ann.jussieu.fr/~frey/cours/UdC/ma691/ma691_ch6.pdf

http://en.wikipedia.org/wiki/Finite_difference_method

http://en.wikipedia.org/wiki/Computational_electromagnetics

http://en.wikipedia.org/wiki/Monte_Carlo_method